# Attosecond correlated electron dynamics at $C_{60}$ giant plasmon resonance

Shubhadeep Biswas[1,2], Andrea Trabattoni[3], Philipp Rupp[1,2], Maia Magrakvelidze[4,5], Mohamed El-Amine Madjet[5,6], Umberto De Giovannini[7], Mattea C. Castrovilli[8,9], Mara Galli[3,10], Qingcao Liu[1,2], Erik P. Månsson[3], Johannes Schötz[1,2], Vincent Wanie[3], François Légaré[11], Pawel Wnuk[1,2], Mauro Nisoli[8,10], Angel Rubio[7,12], Himadri S. Chakraborty[5], Matthias F. Kling[1,2,13,14], Francesca Calegari[3,8,15,16]

[1] Max Planck Institute of Quantum Optics, Hans-Kopfermann-Straße 1, D-85748 Garching, Germany
[2] Department of Physics, Ludwig-Maximilians-Universität München, Am Coulombwall 1, D-85748 Garching, Germany
[3] Center for Free-Electron Laser Science CFEL, Deutsches Elektronen-Synchrotron DESY, Notkestr. 85, 22607 Hamburg, Germany
[4] Science Department, Cabrini University, Radnor, Pennsylvania 19087, USA
[5] Department of Natural Sciences, D.L. Hubbard Center for Innovation, Northwest Missouri State University, Maryville, Missouri 64468, USA
[6] Bremen Center for Computational Materials Science, University of Bremen, Bremen, Germany
[7] Max Planck Institute for the Structure and Dynamics of Matter, Luruper Chaussee 149, D-22761 Hamburg, Germany
[8] Institute for Photonics and Nanotechnologies CNR-IFN, P.za Leonardo da Vinci 32, 20133 Milano, Italy
[9] Istituto Struttura della Materia, ISM-CNR, 00016 Monterotondo Scalo, Roma, Italy
[10] Department of Physics, Politecnico di Milano, Piazza Leonardo da Vinci 32, 20133 Milano, Italy
[11] Institut National de la Recherche Scientifique, 1650 Blvd. Lionel-Boulet, J3X 1S2, Varennes (Qc), Canada
[12] Center for Computational Quantum Physics (CCQ), The Flatiron Institute, New York, NY, USA
[13] SLAC National Laboratory, 2575 Sand Hill Rd, Menlo Park, CA 94025, USA
[14] Applied Physics Department, Stanford University, 348 Via Pueblo, Stanford, CA 94305, USA
[15] The Hamburg Centre for Ultrafast Imaging, Universität Hamburg, 149 Luruper Chaussee, 22761 Hamburg, Germany
[16] Physics Department, University of Hamburg, Luruper Chaussee 149, 22761 Hamburg, Germany

**Fullerenes have unique physical and chemical properties that are associated with their delocalized conjugated electronic structure[1-4]. Among them, there is a giant ultra-broadband - and therefore ultrafast - plasmon resonance, which for $C_{60}$ is in the extreme-ultraviolet energy range[5,6]. While this peculiar resonance has attracted considerable interest for the potential downscaling of nanoplasmonic applications such as sensing, drug delivery and photocatalysis at the atomic-level, its electronic character has remained elusive[7,8]. The ultrafast decay time of this collective excitation demands attosecond techniques for real-time access to the photoinduced dynamics. Here, we uncover the role of electron correlations in the giant plasmon resonance of $C_{60}$ by employing attosecond photoemission chronoscopy. We find a characteristic photoemission delay of up to 200 attoseconds pertaining to the plasmon that is purely induced by coherent large-scale correlations. This result provides novel insight into the quantum nature of plasmonic resonances, and sets a benchmark for advancing nanoplasmonic applications.**

Since their discovery in 1985[1], fullerenes have attracted much interest owing to their unique electronic structure. The most prominent $C_{60}$ fullerene consists of a near-spherical shell of carbon atoms with a diameter of about 0.7 nm and a nearly delocalized cloud of 240 valence electrons around its carbon skeleton (see Fig. 1a). The ultrafast dynamics that characterize these delocalized electrons bring about unique functionalities, in which quantum correlations have been demonstrated to play a major role: for instance, in $C_{60}$-based molecular solids electron correlations can mediate the appearance of narrow bands and the occurrence of high-temperature equilibrium superconductivity[2,3]. Electron correlations are also responsible for the existence of stable $C_{60}$ anions employed as acceptors in organic solar cells[4].

Another important facet of fullerenes is their plasmonic behaviour. The peculiar electronic structure of $C_{60}$, and in particular the coherent dynamics of $\pi$ and $\sigma$ band states resulting from photoexcitation (see Fig. 1a), gives rise to a giant plasmon resonance (GPR). This resonance has a high excitation energy of about 20 eV, well above the ionisation threshold (7.6 eV), and an ultrabroad bandwidth exceeding 10 eV[5,6]. After its prediction and experimental observation in the 1990s[5,6], the GPR in $C_{60}$ has been the subject of intensive studies. Because the GPR engenders an exceptional confinement of light, it offers a unique opportunity of further downscaling the concepts and applications of nanoplasmonics[7] to the single-molecular – or even atomic – level[8]. Despite such tremendous efforts, the exact nature of the GPR lacks a proper understanding, more specifically, in disentangling the role of many-electron quantum-correlation-driven excitations from single-electron excitations. The appearance of a GPR-type resonance relates to geometric confinement, whereas its broadening may involve large-scale coherent electron-electron interaction and scattering. Theoretically, classical models mimicking the collective electron motion around the $C_{60}$ cage have been able to reproduce the excitation energy of the resonance, but not its ultrabroad bandwidth[9] (see Fig. 1b). Quantum theories, instead, while better matching the experimentally observed resonance shape, intrinsically imply the interplay between incoherent single-electron excitations and coherent electron-correlation driven dynamics[10,11]. Experimentally, the GPR has been studied by differential absorption or ionization cross-section measurements and with angle-resolved photoelectron spectroscopy[6,12]. While such studies provided knowledge mostly about the spectral features of the resonance, they don't allow incoherent electron motion to be disentangled from correlation-driven dynamics[11]. As a result, the electronic character of the GPR in $C_{60}$ and, in particular, the role of correlations in the plasmonic response are still an open question. A time-resolved study with attosecond resolution can resolve this question, allowing for the electron motion at the GPR to be tracked in real-time.

The multi-electron character of the $C_{60}$ GPR has been predicted to be encoded in the electronic photoemission delay[13]. In the framework of quantum mechanics, photoemission can intuitively be understood within half-scattering theory: during photoionization the outgoing electron scatters on the surrounding static and time-dependent correlation-induced potential. Each effect induces a phase shift in the electron emission amplitude, all of which are additive[14]. The derivative of the overall phase shift with respect to energy can be associated with a photoemission time delay, which is usually referred to as the Eisenbud–Wigner–Smith (EWS) delay[15-17]. For the case of the GPR in $C_{60}$, the excited plasmon is expected to reshape the potential experienced by the outgoing electron, therefore affecting the total EWS delay. In particular, the presence of electronic correlations, and their primary role in the

GPR, can be demonstrated by measuring the EWS delay with attosecond precision[17,18]. In this context, attosecond spectroscopy enabled real-time tracking of photoemission delays in complex systems[19-21], as well as for phases associated with Fano resonances in atoms[22,23].

In this work, we time-resolve the photoemission from $C_{60}$ upon resonant photo-excitation of the GPR. The photoemission delay was measured for a broad range of detected electron energies around the plasmon excitation. We identified a general energy dependence of the delay for $C_{60}$ that differs significantly from the one of the reference atomic target (neon). Comparing the experimental results with semi-classical and quantum ab-initio models we capture the role of coherent correlated delays solely associated with the GPR dynamics. In particular, we retrieved a photoemission delay of up to 200 as that originates from purely collective dynamics.

The experimental approach employing attosecond streaking spectroscopy[24,25] is schematically illustrated in Fig. 2(a) and described in detail in the Methods. In brief, an isolated extreme-ultraviolet (XUV) attosecond pump pulse was interferometrically combined with a carrier-envelope-phase (CEP) stabilized sub-5 femtosecond near-infrared (NIR) probe pulse[25]. The two synchronized pulses were then focused at variable delay (pulse delay) onto a gas-phase target of isolated $C_{60}$ molecules.

The $C_{60}$ molecules were photoionized by the XUV attosecond pulse in the co-presence of the NIR pulse. The emitted photoelectrons were collected in a velocity map imaging (VMI) spectrometer as a function of the delay between XUV pump and NIR probe[25]. By integrating the acquired photoelectron 2D momentum distributions over a small angle along the laser polarization axis, we obtain the photoelectron kinetic energy distributions for each delay step, resulting in the $C_{60}$ streaking spectrogram shown in Fig. 2b. A second spectrogram (c.f. Fig. 2c) was acquired for neon in the same way (see Methods for details about the synchronization of the two spectrograms) and used as relative timing reference. Aside from the effect of the NIR vector potential giving rise to the overall shape of the spectrogram, energy-dependent temporal shifts are known to be induced by the EWS delay[17,26], the effect of the Coulomb-laser coupling (CLC)[27] and the chirp of the XUV attosecond pulse[25]. For $C_{60}$, additionally, a supplementary phase is induced by the GPR itself, and another by the presence of a NIR-induced dipolar near-field. The overall temporal phase, defined as the streaking delay ($\tau_s$), can be extracted for $C_{60}$ and neon directly from the experimental spectrograms, by fitting isocontour lines for different photoelectron energies. In Figs. 2(b) and (c), representative isocontour lines are shown at two different photoelectron energies, indicating their relative $\tau_s$ (see Methods for further details about the procedure to determine the streaking delays).

The streaking delays as a function of photoelectron energy for the two targets are shown in Fig. 2d. The two curves display opposite trends, with the one of neon[26,27] showing a typical trend expected from the CLC contribution. We note that the extracted streaking delays are defined up to an arbitrary offset that can be calibrated for the reference atomic target (see

Methods for details). Due to the intrinsic synchronization between the measurements on $C_{60}$ and neon, however, this offset is identical for both targets.

In the following discussion, we uncover the contribution of the large-scale electron correlations in $C_{60}$ GPR. For this, we consider relative streaking delays ($C_{60}$ − neon) between $C_{60}$ and neon, which are free from any arbitrary delay offset. Due to the intrinsic synchronization between the measurements for both targets, the relative delay can be obtained with high accuracy and repeatability. It facilitates the comparison of the experimental findings with theoretical predictions. The photoelectron energy dependence of the relative streaking delays is displayed in Fig. 3a, where theoretical predictions are also shown for comparison.

The theoretical simulations combine ab-initio linear-response time-dependent density functional theory (LR-TDDFT)[10] with classical trajectory Monte-Carlo (CTMC) simulations, see Methods for details. The LR-TDDFT theory is used to describe the time dependent response of the electron density to the incident resonant oscillating field in a linear-response frame. The frequency dependent change in the electron density induces a complex potential that accounts for electron correlations. The broad coherent part of LR-TDDFT result is then translated into streaking delays by classical propagation simulations. This propagation part of the simulations particularly incorporates the chirp of the attosecond pulse and the CLC to the streaking delay (see Extended Fig. 1(a)), as found in previous works[25].

In addition, we have used a simple model based on classical electromagnetic theory (see Methods for details) for an intuitive understanding of the role of the fullerene's polarizability[9]. Indeed, the high polarizability of $C_{60}$ results in a coherent oscillation of the electron cloud upon interaction with the NIR field, and, consequently, an oscillating near field with an asymptotic dipolar behaviour. This near-field has an impact on the final phase of the propagating electron that was fully accounted for in the CTMC simulations (see Methods and see Extended Fig. 1(a) for further details). We note that similar results for the near-field induced streaking delay are obtained from the near-field constructed with both classical[9] and quantum density functional theory (DFT)[28] (see Extended Fig. 2 for detailed comparison).

The energy-dependent relative streaking delays obtained from combined LR-TDDFT and CTMC simulations are shown Fig. 3a (purple curve). The experimental relative streaking delays are in excellent agreement with the simulated curve. We note that the residual mismatch between simulation and experiment in the energy range around 13 -15 eV likely originates from the disagreement between LR-TDDFT and experimental cross sections around 20 eV photon energy (see Fig. 1b).

Motivated by the quantitative representation of the experimental data of the simulations, we also compared the data to simulations performed within linear-response density functional theory (LR-DFT)[10] (green curve in Fig. 3(a)). These LR-DFT simulations entirely omit electron correlations, and exclusively inform about the static mean-field scattering delay. By comparing the LR-TDDFT and LR-DFT simulations with the experimental results, the relative contribution of electron-correlation-driven collective versus the non-resonant mean-field dynamics can be uncovered. For the concerned energy range, there is a substantial relative streaking delay difference between the correlated and the mean-field emission cases. This difference increases in the lower photoelectron energy region.

The contribution of the pure plasmonic correlation can be easily quantified by subtracting the LR-DFT curve from that of the LR-TDDFT, as shown in Fig. 3b. Since all propagation effects are identical in both cases, the result can directly be visualized in terms of the EWS delay instead of a measured streaking delay. The resulting curve represents the correlation-driven photoemission delay, i.e. the EWS delay originating exclusively from the large-scale correlation-induced collective excitation of the GPR. It ranges from a minimum of 50 attosecond to about 200 attosecond in the lower energy region.

The LR-TDDFT model also explains the fundamental origin of this large delay. Indeed, the calculations show that a large-scale electron correlation causes an attractive, broad local minimum in the complex induced-potential near the GPR energy (c.f. inset of Fig. 3(b)). Therefore, we may conclude that during its birth process, the outgoing photoelectron scatters through this attractive potential resulting in a transient trapping and a positive delay in its motion.

In conclusion, we employed attosecond photoemission chronoscopy to disclose the role of quantum mechanical electron correlations at the giant plasmon resonance of $C_{60}$. With the support of combined quantum and classical simulations, we demonstrated that the plasmon-activated coherent electron dynamics leads to positive delays to the photoionization process up to 200 attosecond. The excellent agreement of the data with LR-TDDFT theory suggests the reliability of the theoretical model to capture the essence of dominant plasmon dynamics of a finite size electron gas. Our study sets a benchmark for a deeper understanding of plasmonic dynamics in nanomaterials. Given the importance of collective excitations in many fields of science and technology, the study can inspire investigations of rapid decoherence and control of plasmon phenomena on their natural time scale. This will be a crucial step towards informed applications of such systems in quantum nanoplasmonics.

**ACKNOWLEDGEMENT**

F. C. acknowledges support from the German Research Foundation (DFG)—SFB-925—project 170620586, the Cluster of Excellence Advanced Imaging of Matter (AIM) and the European COST Action CA18222 AttoChem. We acknowledge support by the DFG via Kl-1439/10-1 (S.B. and M.F.K.) and LMU excellent (S.B., P.R., Q.L., J.S., P.W., and M.F.K.). We are grateful for support by the Max Planck Society via the IMPRS for Advanced Photon Science (J.S.) and the Max-Planck Fellow program (M.F.K.). H.S.C. acknowledges support from the US National Science Foundation via Grant Nos. PHY-1806206 and PHY-2135107. M.F.K.'s work at SLAC is supported by the U.S. Department of Energy, Office of Science, Basic Energy Sciences, Scientific User Facilities Division, under Contract No. DE-AC02-76SF00515.

**AUTHOR CONTRIBUTIONS**

S.B., A.T., and P.R. contributed equally to this study. M.F.K., H.S.C., and F.C. conceived the project. M.F.K and F.C. coordinated the work. S.B., P.R., M.C.C., M.G., Q.L., E.P.M., J.S., V.W., P.W., and F.C. carried out the experiments. S.B., A.T., and P.R. analyzed the experimental data. M.M., M.E.M., U.D., A.R., and H.S.C. performed the quantum simulations. S.B. performed the classical EM theory simulations for near field calculation. S.B. and P.R. carried out classical Monte-Carlo trajectory simulations. S.B., A.T., H.S.C., M.F.K., and F.C. drafted the manuscript. All authors contributed to the discussion of the results and the editing of the manuscript.

**DATA AVAILABILITY**

The data that support the findings of this study are available from the corresponding author upon reasonable request.

**CODE AVAILABILITY**

The code used for the simulations contained in this study is available from the corresponding author upon reasonable request.

**COMPETING INTERESTS**

The authors declare no competing interests.

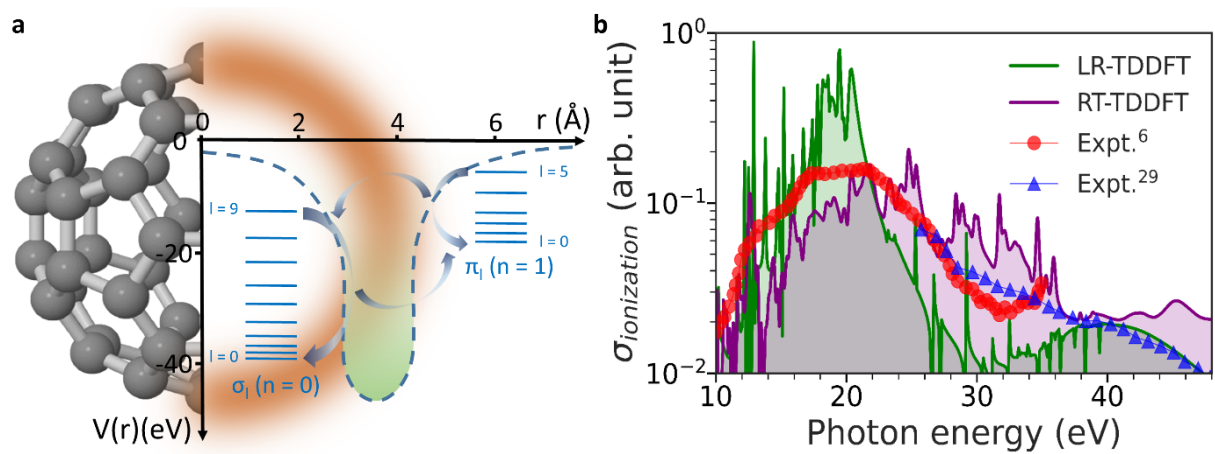

**Figure 1: Electron correlation induced giant plasmon resonance (GPR) in $C_{60}$.** (a) The spherical-shell like distribution of delocalized electrons around $C_{60}$ exhibits a collective giant plasmon excitation at around 20 eV. The jellium-based DFT potential depicted as a function of radial coordinate of $C_{60}$ provides the energetics of the involved quantum states which can be classified into π and σ characters. The configuration interactions (schematically represented by the arrows) among the electrons occupying these states give rise to the GPR. (b) The photoionization cross section as a function of photon energy. The solid curves represent results from TDDFT (LR: linear response; RT: real time) calculations used in this work. These are compared with the experimental photoionization cross sections obtained from Ref. [6] and [29].

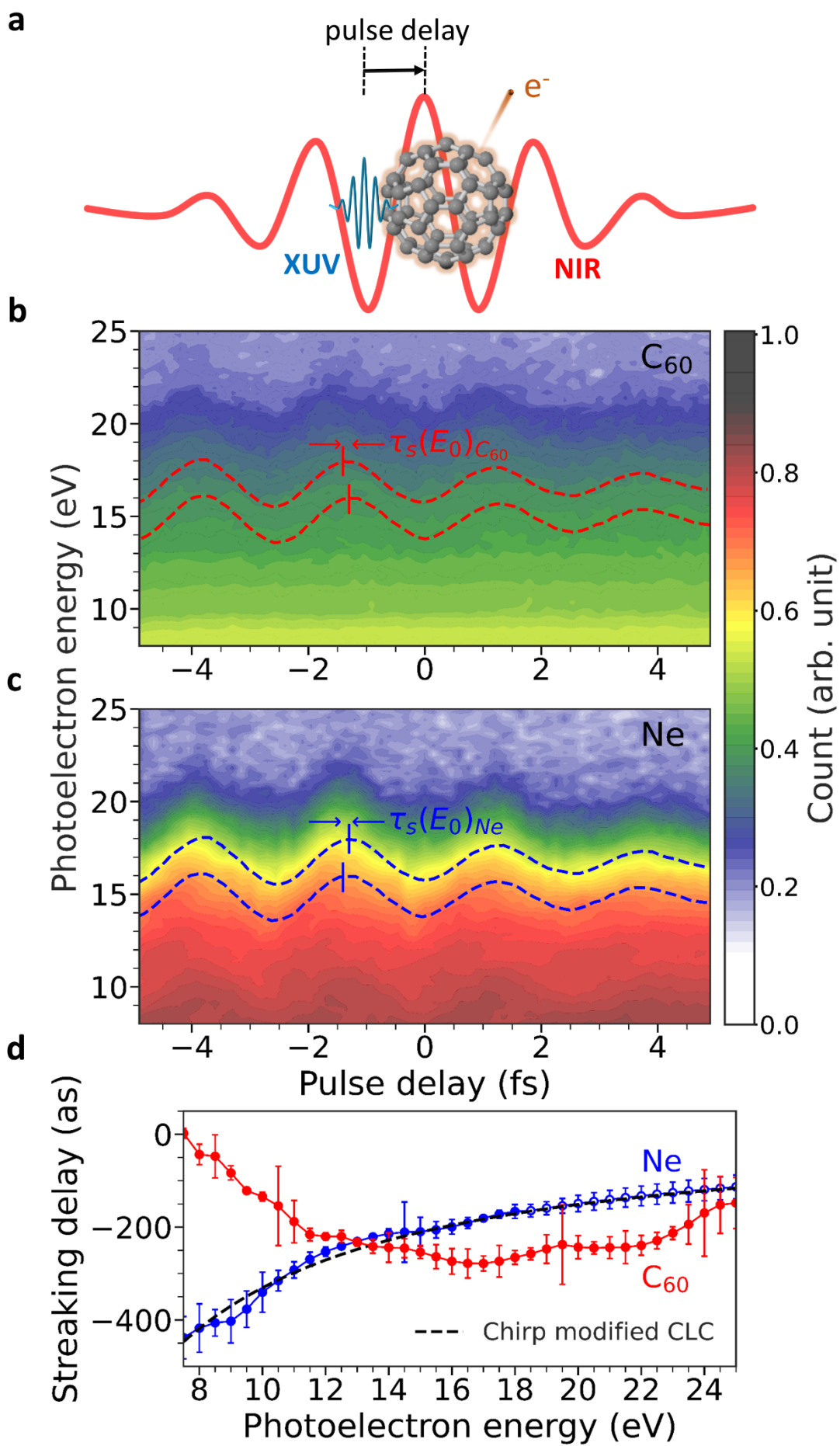

**Figure 2: Attosecond streaking.** (a) For attosecond streaking spectroscopy variably delayed XUV and NIR pulses were employed for ionizing the $C_{60}$ around the GPR and clocking the emission timing, respectively. The photoelectrons are captured and energy analysed by the VMI spectrometer, which results in the streaking spectrogram. (b) and (c) Experimentally measured streaking spectrograms for $C_{60}$ and neon (2p), respectively. The representative streaking curves extracted at given photoelectron energies are shown, whose relative phase difference translates in to the photoelectron energy dependent streaking delays ($\tau_s(E_0)$). (d) Extracted streaking delays as function of photoelectron energy for $C_{60}$ and neon (2p). These results are obtained by weighted averaging over the results produced by five independent measurements, where each measurement also contributes two separate data sets corresponding to two opposite directions along the laser polarization axis. The streaking delay curves share a common time zero reference, which, however, is unknown. The error bars represent the weighted standard deviation including all the measurements. The open symbols for neon results in the higher photoelectron energy side (after 18 eV) indicate the data extracted through extrapolation (see Methods for details). The chirp modified CLC contribution for neon is shown as reference.

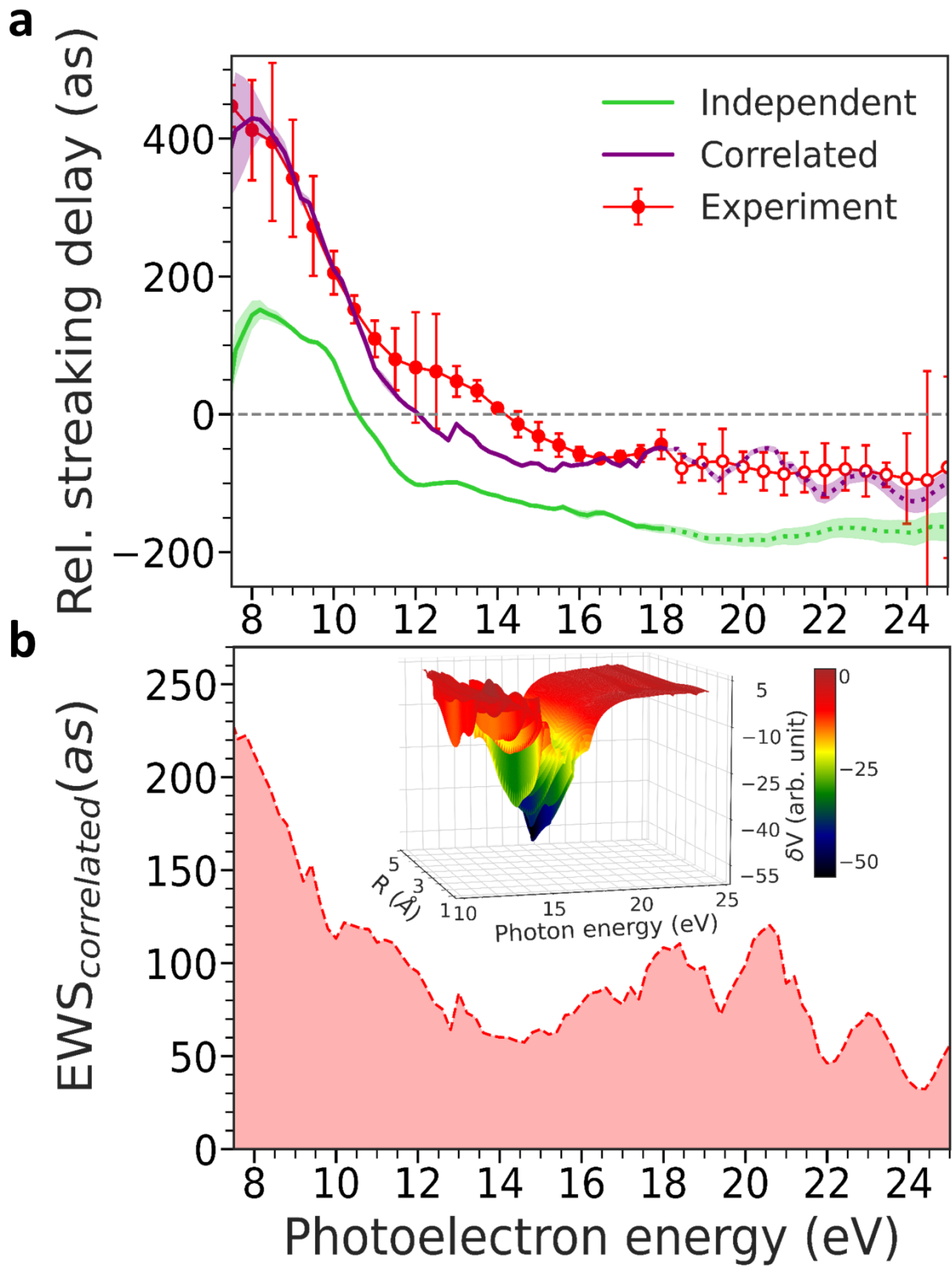

**Figure 3: Relative streaking delay and effect of the GPR on photoemission delay.** (a) Experimentally measured relative streaking delay for $C_{60}$ in comparison to that for neon 2p emission. This data is obtained by weighted averaging over all the relative streaking delays obtained from each pair of $C_{60}$ and neon measurements. The experimental results are compared with the theoretical

simulations which comprise of quantum mechanical DFT calculations and CTMC calculations. The CTMC simulations yield relative streaking delay contributions from XUV chirp, CLC and induced near field effects. The XUV response causing EWS delay is modelled by LR-TDDFT, which takes into account the collective excitation response. For comparison, the independent particle (mean-field) response is modelled by LR-DFT. To compare these with the experimental data in the relative streaking delay level, the CTMC results are added to the EWS results yielding the purple and the green curves representing collective and independent responses, respectively. The error bars in the experimental curve represent the weighted standard deviation corresponding to all the above-mentioned measurements. The shaded regions in the theoretically simulated curves represent the confidence intervals. The open symbols for experimental data and the dotted-line portions in the LR-DFT and LR-TDDFT results in the higher photoelectron energy side (after 18 eV) indicate the data extracted through extrapolation (see Methods for details). (b) The EWS delay contribution exclusively from the correlated excitation is extracted from the difference between the results of LR-TDDFT (correlated) and LR-DFT (mean-field) calculations. The inset represents the induced potential due to GPR. The attractive well-like structure in it causes the transient trapping of the scattered electrons.